\pdfoutput=1

\documentclass[11pt]{article}

\usepackage[preprint]{acl}

\usepackage{times}
\usepackage{latexsym}

\usepackage[T1]{fontenc}

\usepackage[utf8]{inputenc}

\usepackage{microtype}

\usepackage{inconsolata}

\usepackage{graphicx}

\usepackage{booktabs}
\usepackage{threeparttable} 
\usepackage{makecell}
\usepackage{array}
\usepackage{multirow}
\usepackage{multicol}
\usepackage{subfigure}
\usepackage{amsmath}
\usepackage[ruled,linesnumbered,vlined]{algorithm2e}
\usepackage{bm}
\usepackage{tcolorbox}

\title{Enhancing Recommendation Explanations through\\ User-Centric Refinement}

\author{Jingsen Zhang, Zihang Tian, Xueyang Feng, Xu Chen\thanks{Corresponding Author.} \\
        Renmin University of China \\
        \texttt{\{zhangjingsen,xu.chen\}@ruc.edu.cn} \\
        }

\begin{document}
\maketitle
\begin{abstract}
Generating natural language explanations for recommendations has become increasingly important in recommender systems. 
Traditional approaches typically treat user reviews as ground truth for explanations and focus on improving review prediction accuracy by designing various model architectures. 
However, due to limitations in data scale and model capability, these explanations often fail to meet key user-centric aspects such as factuality, personalization, and sentiment coherence, significantly reducing their overall helpfulness to users. 
In this paper, we propose a novel paradigm that refines initial explanations generated by existing explainable recommender models during the inference stage to enhance their quality in multiple aspects.
Specifically, we introduce a multi-agent collaborative refinement framework based on large language models.
To ensure alignment between the refinement process and user demands, we employ a plan-then-refine pattern to perform targeted modifications.
To enable continuous improvements, we design a hierarchical reflection mechanism that provides feedback on the refinement process from both strategic and content perspectives.
Extensive experiments on three datasets demonstrate the effectiveness of our framework.
\end{abstract}

\section{Introduction}
\label{sec:introduction}

Natural language explainable recommendation (NLER) aims to provide users with textual explanations that clarify why an item is recommended or not.
Due to its flexibility and easy of understanding, NLER has become a prominent approach in explainable recommendation~\cite{zhang2020explainable}.
In this field, researchers usually use user reviews as ground truth for explanations and focus on developing advanced architectures to improve review prediction accuracy.
For instance,
NETE~\cite{li2020generate} incorporates specific features into the generation process.
PETER~\cite{li2021personalized} produces explanation by bridging IDs and texts into Transformer. 
PEPLER~\cite{li2023personalized} leverages prompt learning to further enhance the explanation quality.

\begin{figure}[t]
	\centering
	\setlength{\fboxrule}{0.pt}
	\setlength{\fboxsep}{0.pt}
	\fbox{
		\includegraphics[width=0.99\linewidth]{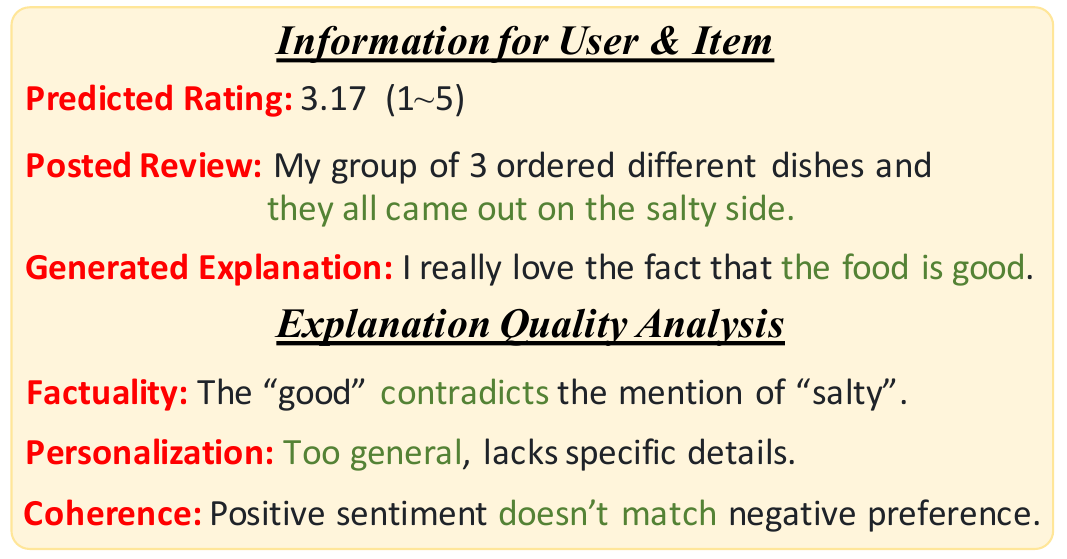}
	}
	\vspace{-0.5cm}
	\caption{Illustration of inadequate user-centric quality in explanations produced by the PETER model.}
	\vspace{-0.5cm}
	\label{fig:intro}
\end{figure}

While leveraging reviews can help explanations partially capture user preference, limitations in data scale and model capability often hinder the overall helpfulness of these explanations, resulting in deficiencies in key user-centric aspects.
To effectively support user decision-making, it is essential for explanations to align with user demands on various aspects.  
For example, factuality ensures that the content is correct and verifiable, personalization requires explanations to highlight specific item features and user characteristics, and coherence demands alignment between the explanation's sentiment and the system-predicted user preference.
Figure~\ref{fig:intro} illustrates these shortcomings, underscoring the importance of generating explanations that are both accurate and genuinely helpful.  

Inspired by OpenAI o1, which enhances model reasoning ability through stepwise thinking during inference and refines its strategies before responding, 
we propose a novel paradigm to refine the initial explanations generated by existing explainable models, achieving targeted improvements in multiple user-centric aspects. 
This represents the first attempt in the field of explainable recommendation to modify explanations during the inference stage before presenting them to users. 
Similar to the Reranking process in recommender systems~\cite{pei2019personalized,gao2024llm}, our paradigm enhances performance in a post-hoc manner, which we define as Refinement.
Leveraging the powerful natural language processing capabilities, we employ large language models (LLMs) to perform refinement with carefully crafted instructions.

While this idea seems promising, it faces several challenges.
First, user demands for explanations are usually multifaceted, and various aspects may hold different priorities~\cite{zhang2024large,rahdari2024logic}. 
Thus, our approach should effectively align with user demands and perform targeted refinements.
Second, existing explainable recommender models~\cite{cao2021cliff,cheng2023explainable,raczynski2023problem} typically provide explanations in a single attempt and lack feedback on whether they meet user demands, often failing to satisfy users. 
Therefore, it is crucial to obtain feedback from the refinement process and enable explanations to evolve from weak to strong, achieving continuous improvements.

To address these challenges, we propose an LLM-based multi-agent collaborative framework for explanation refinement.
Specifically, it adopts a plan-then-refine pattern for targeted modifications guided by user demands, where the Planner first identifies which aspect should be refined at each round, and then the Refiner modifies the explanations according to the corresponding instructions.
Additionally, to ensure continuous improvement, we design a hierarchical reflection mechanism, where the Reflector provides timely feedback and suggestions by analyzing the refinement process from both strategic and content perspectives.
Furthermore, to support this process, we maintain an aspect library containing essential information about various aspects.
By combining forward refinement and backward reflection phases, our framework achieves self-evolving and iterative enhancement until user demands are satisfied.

The main contributions of this paper are as follows:
(1) We identify key limitations of existing explainable recommender models in user-centric aspects and propose a novel approach to perform targeted \underline{Refine}ment to e\underline{X}planations (called \textbf{RefineX} for short).
To our knowledge, it is the first time in the field of explainable recommendation.
(2) We design an LLM-based multi-agent collaborative refinement framework, which employs a plan-then-refine pattern to align with user demands and incorporates a hierarchical reflection mechanism for continuous improvement. 
(3) Extensive experiments demonstrate our framework's high adaptability to diverse user demands and its effectiveness in enhancing explanation quality.


\section{Preliminary}
\label{sec:preliminary}
\textbf{Natural Language Explainable Recommendation (NLER)} aims to provide textual explanations clarifying why an item is recommended or not.
Formally, given a user set $\mathcal{U}$ and an item set $\mathcal{I}$, their interactions are recorded in $\mathcal{D}=\{u, i, r_{u,i}, \bm{s}_{u,i} | u \in \mathcal{U}, i \in \mathcal{I}\}$, where $r_{u,i}$ denotes the user's rating to the item (ranging from 1 to 5), and $\bm{s}_{u,i}$ represents the user's review, which is typically treated as the ground truth for explanation in existing research~\cite{li2021personalized, cheng2023explainable, zhou2024enhancing}.
For a user-item pair $(u,i)$, the goal of NLER is not only to predict the rating $\hat{r}_{ui}$ indicating the user's preference for $i$, but also to generate a textual explanation $\bm{e}_{ui}$ that justifies the recommendation results.
Since rating prediction has been well studied in recommender systems, this work focuses on the explanation generation task.
In this field, metrics such as BLEU~\cite{papineni2002bleu} and ROUGE~\cite{lin2004rouge} are commonly used to evaluate model performance by measuring the similarity between generated explanations and user reviews. 
However, approaches that directly predict user reviews struggle to meet user-centric aspects, limiting the helpfulness of these explanations.

\textbf{Task Formulation.}
In real-world recommendation scenarios, user demands for explanations are often multifaceted and focus on practical helpfulness~\cite{zhang2024large,rahdari2024logic}.
Our task is to refine the explanations generated by existing models to elevate their quality on user demands.
Formally, given a user-item pair $(u,i)$, the user demand (or goal) for explanations are denoted as $G_{ui}=\{ a_{1}, a_{2}, ..., a_{n} \}$, where each $a\in G_{ui}$ represents a specific user-concerned aspect.
Guided by the user goal, our task is to iteratively refine the initial explanation $\bm{e}_{ui}^{0}$ over $t$ rounds to obtain the final explanation $\bm{e}_{ui}^{t}$, achieving the objective:
\begin{equation}
    Q_{a}(\bm{e}_{ui}^{t})>Q_{a}(\bm{e}_{ui}^{0}), \forall a \in G_{ui},
    \label{equ:objective}
\end{equation}
where $Q_{a}$ is the evaluation function that measures the explanation quality on the aspect $a$. 
The key challenge in this task is aligning refinements with the user goal and providing feedback in each round to achieve continuous improvement.


\begin{figure*}[t]
	\centering
	\setlength{\fboxrule}{0.pt}
	\setlength{\fboxsep}{0.pt}
	\fbox{
		\includegraphics[width=0.92\linewidth]{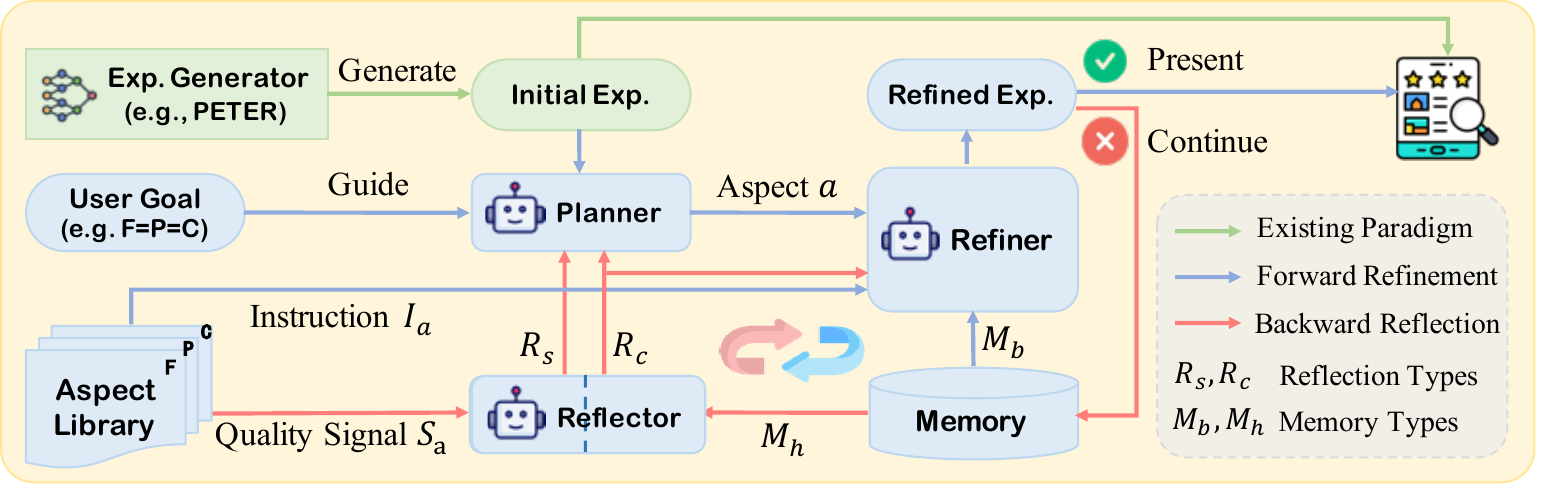}
	}
	\vspace{-0.1cm}
	\caption{The overview of our RefineX framework, where ``Exp'' denotes ``Explanation''.}
	\vspace{-0.3cm}
	\label{fig:model}
\end{figure*}

\section{Approach}
\label{sec:approach}
In this section, we introduce RefineX, an LLM-based multi-agent collaborative framework designed to refine recommendation explanations and enhance their quality in user-centric aspects.

\subsection{Framework Overview}
The overall framework is shown in Figure~\ref{fig:model}.
Traditional NLER models directly provide users with explanations without feedback on whether they meet user goals.
Our framework addresses this gap by refining generated explanations based on user goals before presenting them to users.
Specifically, the entire process consists of a forward refinement phase and a backward reflection phase.
In the refinement phase, we adopt a plan-then-refine pattern, where the Planner first identifies the aspects needing refinement, and then the Refiner performs targeted modifications.
In the reflection phase, the Reflector analyzes the behaviors of the Planner and the Refiner from both strategic and content perspectives and provides feedback for improvement. 
Additionally, to support this process, we construct a Memory module and Aspect Library to maintain essential information about the refinement.
We detail these components below.

\subsection{The Refinement Phase}
LLMs have demonstrated powerful planning capabilities~\cite{huang2022language,wang2023describe,huang2022inner} in handling complex tasks. 
To align with users' multifaceted goals and perform targeted refinements, we adopt a plan-then-refine pattern for each round of the refinement.
This phase involves the following components:

\textbf{Planner} plays a crucial role in controlling the refinement process, determining which aspect to refine in each round or terminating the process when the explanation meets user goals.
We implement the Planner's behavior using the ReAct~\cite{yao2022react} framework, which employs a thought-action-observation pattern for task solving.
Specifically, for round $t$, the Planner first determines which aspect $a^{t}$ of the previous explanation $\bm{e}^{t-1}$ to refine by analyzing the user goal $G$, refinement trajectory $\bm{T}^{1:t}$, and reflections from the previous round $t-1$ at the strategic level $R_{s}^{t-1}$ and content level $R_{c}^{t-1}$ (details in Section~\ref{sec: reflection}).
The Planner's thought function is formulated as:
\begin{equation}
    a^{t} = \text{Planner}(\bm{e}^{t-1}, G, \bm{T}^{1:t-1}, R_{s}^{t-1}, R_{c}^{t-1}),
    \label{equ: planner}
\end{equation}
where $\bm{T}^{1:t}=[a^{1}, a^{2}, ..., a^{t}]$ represents the trajectory of refined aspects, and the subscript ``$ui$'' for the specific user-item pair $(u, i)$ is omitted for clarity. 
Once the refined aspect is determined, the Planner takes action to call the corresponding information acquisition functions organized in the aspect library $\mathcal{A}$ (details in Appendix~\ref{sec:aspect library}), and the obtained aspect instructions for the sample $(u, i)$ are conveyed to the Refiner for targeted refinement.

\textbf{Refiner} is the dedicated agent responsible for refining the previous explanation $\bm{e}^{t-1}$ on the aspect $a^{t}$ selected by the Planner.
This process is guided by the refinement instructions $I_{a^{t}}$ for aspect $a^{t}$, including standards and auxiliary information (\textit{e.g.}, item characteristics).
Additionally, reflections on refined explanation content ${R}_{c}^{1:t-1}$ from the past rounds are also used to guide the Refiner's refinement performance. 
To ensure relevance, we summarize the reflections about aspect $a^{t}$, following prior works~\cite{zhang2024survey, wang2023user}, denoted as:  
\begin{equation}
    \bm{\tilde{R}}_{c}^{1:t-1} = \text{Summarize}(a^{t}, \bm{R}_{c}^{1:t-1}) 
    \label{equ: summary}
\end{equation}
The Refiner's refinement function is formulated as: 
\begin{equation}
    \bm{e}^{t} = \text{Refiner}(\bm{e}^{t-1}, a^{t}, I_{a^{t}}, \bm{\tilde{R}}_{c}^{1:t-1}) 
    \label{equ: refiner}
\end{equation}

\textbf{Memory} module is designed to store key information about the target sample and record the refinement process. 
It consists of two components: Background Memory and Refinement Memory.
Background memory $M_{b}$ includes profiles of the target user-item pair $(u, i)$, such as item attributes (\textit{e.g.}, title and category) and user interactions (\textit{e.g.}, ratings and reviews).
This component serves as the resource for acquiring auxiliary information using acquisition functions in the aspect library.
Refinement memory $M_{h}$ records the refinement history, serving as the reference for producing reflections.
The refinement record $h^{t}$ in round $t$ includes the refined explanation $\bm{e}^{t}$, refined aspect $a^{t}$, refinement instructions $I_{a^{t}}$, and two levels of reflection, $R_{s}^{t}$ and $R_{c}^{t}$, denoted as:
\begin{equation} 
    h^{t} = \{ \bm{e}^{t}, a^{t}, I_{a^{t}}, R_{s}^{t}, R_{c}^{t}  \} 
    \label{equ: memory}
\end{equation}
The refinement memory is updated at each round as $\bm{M}_{h}^{1:t} = \bm{M}_{h}^{1:t-1} \cup \{ h^{t} \}$.
This structured memory effectively supports information acquisition and reflection generation.

\subsection{The Reflection Phase}
\label{sec: reflection}
To provide feedback and suggestions to the Planner and Refiner, 
we introduce a hierarchical reflection mechanism to assess the refinement process from both strategic and content perspectives.
This mechanism consists of two components:

\textbf{Strategic Reflector} aims to examine the alignment of the refinement trajectory with user goals through rule-based reflection, focusing on three key criteria: accuracy, completeness, and priority.
Specifically, with the guidance of user goals, it mainly evaluates whether the refinement process involves irrelevant aspects, omits key aspects, or overlooks the relative priority of aspects.
Besides providing feedback on planning, this Reflector is also expected to offer constructive suggestions to enhance the Planner. 
The generation of strategic reflection is formulated as:
\begin{equation}
    R^{t}_{s} = \text{S\_Reflector}(G, \bm{M}_{h}^{1:t}, C_{s}),
    \label{equ: strategic}
\end{equation}
where $G$ is the user goal, $\bm{M}_{h}^{1:t}$ represents the refinement history, and $C_{s}$ denotes the evaluation criteria at the strategic level.

\textbf{Content Reflector} evaluates the refinement performance from the perspective of explanation content. 
It provides insights about whether the explanation conforms to several content criteria, such as following the refinement instructions, covering necessary details, and excluding irrelevant content.
Notably, it achieves tool-augmented reflection by calling external aspect metrics in the aspect library at each round and these values serve as quality reference signals for more comprehensive evaluation. 
The Content Reflector provides a timely updated view of explanation quality to the Planner and the Refiner for targeted planning and refinement. 
Its generation function is:
\begin{equation}
    R^{t}_{c} = \text{C\_Reflector}(\bm{e}^{t}, a^{t}, I_{a^{t}}, S_{a^{t}}, C_{c}),
    \label{equ: content}
\end{equation}
where $I_{a^{t}}$ and $S_{a^{t}}$ are the refinement instructions and external quality signal for aspect $a^{t}$, respectively.
$C_{c}$ denotes the content criteria.

In summary, these two types of Reflectors complement each other in evaluating the refinement process. 
The Strategic Reflector provides high-level feedback, ensuring overall alignment between planning and user goals, while the Content Reflector offers fine-grained feedback, ensuring the precision of refined content on specific aspects.
This hierarchical reflection mechanism enables more aligned and precise enhancement of the refinement process.
Additionally, to support the entire process, we construct an aspect library containing key information on several user-centric aspects.
For more details, please refer to Appendix~\ref{sec:aspect library}.

\subsection{Discussion}

Considering the complex and diverse nature of user behaviors in recommender systems, directly generating explanations using LLMs may struggle to accurately identify user preferences~\cite{lei2024recexplainer, ma2024xrec}, since LLMs are less effective at processing domain-specific data, such as collaborative information in recommender systems.
Instead, we propose refining the explanations generated by existing explainable recommender models, which contain predicted user preference. 
This approach effectively combines the capabilities of recommender models in preference identification with the strengths of LLMs in language generation, ensuring the production of both accurate and helpful explanations.

Furthermore, our proposed multi-agent collaborative refinement framework includes a forward refinement phase for planning and refining tasks, and a backward reflection phase for feedback acquisition to improve the system.
This process is analogous to model optimization in deep learning, where the model performs forward inference to generate task outputs, while gradients are backpropagated to update model parameters.
The complete process of our framework is presented in Algorithm~\ref{pipeline}.

\setlength{\textfloatsep}{0.2cm}
\begin{algorithm}[ht] 
\small
\caption{The Process of Refinement.} 
\label{pipeline} 
Specify the user goal $G$ and the maximum number of refinement rounds $N$.\\
Prepare the aspect library $\mathcal{A}$.\\
Initialize the background memory $M_{b}$.\\
Generate the initial explanation $\bm{e}^{0}$ using the pre-trained explainable recommender model.\\
\For{round $t$ in [1, $N$]}{
	\tcp{The Refinement Phase:}
            Obtain a plan from the Planner using Eq.~(\ref{equ: planner}).\\
            \If{fully refined}{
                Terminate the process.
                }
            \ElseIf{an aspect $a^{t}$ is selected to refine}{
                Retrieve sample information from $M_{b}$ by calling functions in $\mathcal{A}$.\\
                Summarize content reflections for aspect $a^{t}$ using Eq.~(\ref{equ: summary}).\\
                Generate the refined explanation $e^{t}$ by the Refiner using Eq.~(\ref{equ: refiner}).\\
                }
	\tcp{The Reflection Phase:}
            Obtain the strategic reflection $R_{s}^{t}$ using Eq.~(\ref{equ: strategic}).\\
            Derive the external quality signals $S_{a^{t}}$ from $\mathcal{A}$.\\
            Obtain the content reflection $R_{c}^{t}$ using Eq.~(\ref{equ: content}).\\
            \BlankLine
        Update refinement memory $M_{h}$ using Eq.~(\ref{equ: memory}).
}
Output the final explanation to the user.
\end{algorithm}

\section{Experiments}
\label{sec:experiments}

\subsection{Experiment Setup}

\textbf{Datasets.}
We conduct experiments using three real-world datasets from distinct domains.
\textbf{Yelp}~\footnote{https://www.yelp.com/dataset} includes user ratings and reviews of various restaurants.
\textbf{Amazon-Beauty~\footnote{https://jmcauley.ucsd.edu/data/amazon/index\_2014.html} (Beauty)} and \textbf{Amazon-Video Games (Games)} contain user interactions regarding cosmetics and video games, respectively, on the Amazon e-commerce platform.
Detailed statistics of the datasets are presented in Table~\ref{tab:dataset}.

\textbf{Baselines.}
We utilize \textbf{PETER}~\cite{li2021personalized}, a commonly used model in explainable recommendation, as the base model to generate initial explanations.
We compare PETER's performance with two types of enhanced methods:
(1) Model-Oriented methods, which enhance a specific aspect by modifying model architectures or training processes: 
\textbf{CLIFF}~\cite{cao2021cliff} for factuality, 
\textbf{ERRA}~\cite{cheng2023explainable} for personalization, and \textbf{CER}~\cite{raczynski2023problem} for coherence.
(2) Model-Agnostic methods, which refine explanations generated by the base model in a post-hoc manner: 
the LLM-based approach \textbf{LLMX}~\cite{luo2023unlocking} and our approach \textbf{RefineX}.

\textbf{Evaluation Metrics.}
To evaluate the factuality of explanations, we employ \textbf{Entailment Ratio (Entail)}~\cite{xie2023factual, zhuang2024improving}, which assesses the proportion of explanations that can be entailed or supported by existing reviews. 
For personalization, we first use \textbf{Feature Coverage Ratio (FCR)}~\cite{li2020generate, li2021personalized} to evaluate at the feature level by measuring the fraction of features present in the explanations.  
Additionally, we utilize \textbf{ENTR}~\cite{jhamtani2018learning, xie2023factual} to assess personalization from the diversity perspective by calculating the entropy of the n-grams distribution in the generated text.
To evaluate coherence, we apply \textbf{Coherence Ratio (CoR)}~\cite{yang2021explanation, zhao2024aligning}, which measures the proportion of samples achieving coherence between the sentiment of generated explanations and the predicted user preference.

\begin{table*}[t]
    \caption{Performance comparison of various approaches. The best and the second-best performances are highlighted using bold and underlined fonts, respectively. For all metrics, a higher value indicates better performance.}
    \vspace{-0.2cm}
    \center
    \renewcommand\arraystretch{1.1} 
    \setlength{\tabcolsep}{7pt}{
        \begin{threeparttable}  
            \scalebox{0.83}{
                \begin{tabular}
                    {   p{1.2cm}<{\centering}|
                        p{0.9cm}<{\centering}p{0.9cm}<{\centering}p{1.0cm}<{\raggedleft}p{1.0cm}<{\centering}|
                        p{0.9cm}<{\centering}p{0.9cm}<{\centering}p{1.0cm}<{\raggedleft}p{1.0cm}<{\centering}|
                        p{0.9cm}<{\centering}p{0.9cm}<{\centering}p{1.0cm}<{\raggedleft}p{1.0cm}<{\centering}
                    } 
                    \toprule
                    \multirow{2}{*}{\textbf{Method}}&\multicolumn{4}{c|}{\textbf{Yelp}}&\multicolumn{4}{c|}{\textbf{Beauty}}&\multicolumn{4}{c}{\textbf{Games}}\\        
                    \cmidrule(lr){2-5} \cmidrule(lr){6-9} \cmidrule(lr){10-13} 
                    &\textbf{Entail}&\textbf{FCR}&\textbf{ENTR}&\textbf{CoR}&\textbf{Entail}&\textbf{FCR}&\textbf{ENTR}&\textbf{CoR}&\textbf{Entail}&\textbf{FCR}&\textbf{ENTR}&\textbf{CoR}\\ 
                    \midrule
                    PETER & 0.295 & 0.0166 & 7.089 & 0.455 & 0.265 & 0.0286 & 5.637 & 0.605 & 0.380 & 0.0078 & 6.280 & 0.555 \\
                    CLIFF & 0.450 & 0.0091 & 6.062 & 0.445 & 0.430 & 0.0233 & 6.254 & 0.745 & 0.395 & 0.0132 & 6.341 & \underline{0.665} \\
                    CER   & 0.195 & 0.0155 & 6.529 & \underline{0.530} & 0.450 & 0.0318 & 6.444 & \underline{0.840} & 0.300 & 0.0102 & 5.918 & 0.625 \\
                    ERRA  & 0.185 & 0.0178 & 6.922 & 0.470 & 0.220 & 0.0424 & 6.709 & 0.790 & 0.325 & 0.0112 & 6.202 & 0.520 \\
                    \midrule
                    LLMX  & \underline{0.555} & \underline{0.0299} & \underline{9.721} & \underline{0.530} & \underline{0.575} & \underline{0.1081} & \underline{9.459} & 0.765 & \underline{0.670} & \underline{0.0351} & \underline{9.527} & 0.635 \\
                    RefineX & \textbf{0.835} & \textbf{0.0424} & \textbf{10.082} & \textbf{0.770} & \textbf{0.800} & \textbf{0.1441} & \textbf{10.160} & \textbf{0.885} & \textbf{0.835} & \textbf{0.0473} & \textbf{10.099} & \textbf{0.750} \\

                    \bottomrule
                \end{tabular}
            }
    \end{threeparttable}}
    \vspace{-0.2cm}
    \label{tab:overall}   
\end{table*}

\begin{table}[t]
    \caption{Ablation study of the reflection mechanism. ``SR'' and ``CR'' denote ``strategic reflector'' and ``content reflector'', respectively.}
    \vspace{-0.2cm}
    \center
    \renewcommand\arraystretch{1.05} 
    \setlength{\tabcolsep}{6pt}{
        \begin{threeparttable}  
            \scalebox{0.8}{
                \begin{tabular}
                    {   p{1.05cm}<{\raggedright}|
                        p{2.05cm}<{\raggedright}|
                        p{0.9cm}<{\centering}p{0.9cm}<{\centering}p{0.9cm}<{\centering}p{1.0cm}<{\centering}
                    } 
                    \toprule
                    \textbf{Dataset}&\textbf{Method}&\textbf{Entail}&\textbf{FCR}&\textbf{ENTR}&\textbf{CoR}\\ 

                    \midrule
                    \multirow{4}{*}{\textbf{Yelp}}&RefineX & \underline{0.835} & \textbf{0.0424} & \textbf{10.082} & \textbf{0.770} \\
                    &\textit{-w/o} SR&\textbf{0.845}    &\underline{0.0393} &\underline{10.080}&0.715\\
                    &\textit{-w/o} CR&0.820 &0.0392 &10.041&\underline{0.755}\\
                    &\textit{-w/o} SR\&CR&0.760 &0.0363 &10.011&0.700\\

                    \midrule
                    \multirow{4}{*}{\textbf{Beauty}}&RefineX & \textbf{0.800} & \textbf{0.1441} & \textbf{10.160} & \textbf{0.885} \\
                    &\textit{-w/o} SR&0.755 &\underline{0.1292} &\underline{10.111}&\underline{0.855}\\
                    &\textit{-w/o} CR&0.710 &0.1250 &10.041&0.830\\
                    &\textit{-w/o} SR\&CR&\underline{0.760} &0.1239 &10.093&0.820\\

                    \bottomrule
                \end{tabular}
            }
    \end{threeparttable}}
    \vspace{-0.0cm}
    \label{tab:ablation}   
\end{table}

\textbf{Implementation.}
We organize each user's interactions chronologically for all datasets and divide training and testing sets using the \textit{leave-one-out} strategy.
Due to budget constraints, we follow recent LLM-based recommendation studies~\cite{zhang2024agentcf,zhao2024let,huang2023recommender,zhou2024cognitive} and randomly sample 200 users from the testing set of each dataset for evaluation.
We implement two post-hoc methods, LLMX and RefineX, using GPT-3.5~\footnote{gpt-3.5-turbo-0125}, with the user goal defaulted as ``Assign equal importance to three aspects: factuality, personalization and sentiment coherence."
More details on experiment setup are provided in Appendix~\ref{sec:implementation}.
Prompt templates used in our framework are presented in Appendix~\ref{sec:prompt}.

\subsection{Overall Performance}
The overall comparison results between RefineX and baselines are presented in Table~\ref{tab:overall}.
We can see:

(1) Compared to the based model PETER, model-oriented methods involve modifications to architectures in specific aspects, and achieve improvement in the corresponding metrics. 
This is consistent with the findings reported in the original papers: CLIFF performs better on Factuality, ERRA on Personalization, and CER on Coherence.
However, we note that these improvements are limited and unstable. 

(2) In contrast, model-agnostic methods, which refine explanations generated by PETER in a post-hoc manner, achieve more comprehensive improvements across multiple aspects. 
We speculate that user-centric explanations place high demands on both user preference and textual expression.
Based on the predicted user preference by the base model, these methods utilize the powerful LLMs to achieve more natural and fluent expressions, which contributes to superior performance.

(3) Notably, our RefineX framework achieves the best performance, and the superiority is consistent across all datasets and aspects.
These observations verify the effectiveness of our refinement framework in enhancing explanations in user-centric aspects. 
Our framework realizes these through a multi-agent collaborative mechanism, which adopts a plan-then-refine pattern for targeted refinement and employs hierarchical reflections for continuous refinement.

\begin{table*}[t]
    \caption{Adaptability analysis of RefineX on various user goals. 
    The comparison symbols in goals indicate the priority of different aspects: Factuality (F), Personalization (P) and Coherence (C).
    ``Aspect Ratio'' represents the proportion of refined aspects in all testing samples. 
    ``Representative'' and ``Ratio'' denote the most representative refinement trajectory under each goal and its corresponding ratio.
    ``Length'' and ``Max Stop'' refer to the average length of trajectory and the proportion of samples reaching the maximum number of rounds, respectively.
    }
    \vspace{-0.2cm}
    \center
    \renewcommand\arraystretch{1.05} 
    \setlength{\tabcolsep}{7pt}{
        \begin{threeparttable}  
            \scalebox{0.8}{
                \begin{tabular}
                    {   p{1.1cm}<{\raggedright}|
                        p{1.1cm}<{\centering}|
                        p{1.0cm}<{\centering}p{1.0cm}<{\centering}p{1.0cm}<{\centering}p{1.0cm}<{\centering}|
                        p{2.1cm}<{\centering}|
                        p{2.3cm}<{\raggedright}p{0.9cm}<{\raggedright}|
                        p{1.0cm}<{\centering}p{1.6cm}<{\centering}
                    } 
                    \toprule
                    \multirow{2}{*}{\textbf{Dataset}} & \multirow{2}{*}{\textbf{Goal}} & \multicolumn{4}{c|}{\textbf{Explanation Quality}} & \textbf{Aspect Ratio} & \multicolumn{4}{c}{\textbf{Trajectory}}\\
                    \cmidrule(lr){3-6} \cmidrule(lr){7-7} \cmidrule(lr){8-11}
                    & & \textbf{Entail} & \textbf{FCR} & \textbf{ENTR} & \textbf{CoR} & \textbf{F : P : C} & \textbf{Representative} & \textbf{Ratio} & \textbf{Length} & \textbf{Max Stop} \\ 
                    \midrule
                    \multirow{3}{*}{\textbf{Yelp}} 
                    & F=P=C & 0.835 & \textbf{0.0424} & 10.082 & \textbf{0.770} & 32 : 40 : 28 & [F, P, C] & 30.4\% & 4.00 & 34.5\% \\
                    & F>P>C & \textbf{0.855} & 0.0413 & \underline{10.085} & \underline{0.605} & 46 : 38 : 16 & [F, P, C, F] & 10.9\% & 3.87 & 38.0\%\\
                    & P>F>C & 0.790 & \underline{0.0422} & \textbf{10.119} & 0.545 & 16 : 73 : 11 & [P, P, P, F, P, C] & 12.5\% & 4.47 & 50.5\%\\
                    \midrule
                    \multirow{3}{*}{\textbf{Beauty}} 
                    & F=P=C & \textbf{0.800} & \textbf{0.1441} & \underline{10.160} & \textbf{0.885} & 34 : 41 : 25 & [F, P, C] & 35.4\% & 4.06 & 31.5\%\\
                    & F>P>C & \underline{0.790} & 0.1314 & 10.139 & 0.805 & 48 : 35 : 17 & [F, P, C, F] & 16.4\% & 3.96 & 34.5\%\\
                    & P>F>C & 0.705 & \underline{0.1356} & \textbf{10.206} & \underline{0.825} & 18 : 68 : 14 & [P, C, F] & 10.3\% & 4.01 & 39.0\%\\
                    \bottomrule
                \end{tabular}
            }
    \end{threeparttable}}
    \vspace{-0.2cm}
    \label{tab:adaptability}   
\end{table*}

\subsection{Ablation Study of the Reflection}

Reflections provide analyses and suggestions for the refinement process, playing a crucial role in system self-evolution.
In this section, we investigate the impact of various components within our reflection mechanism by sequentially removing strategic and content reflections.
The results from the Yelp and Beauty datasets are presented in Table~\ref{tab:ablation}, revealing the following insights:
Removing each level of reflection leads to a decline in explanation quality.
The worst performance appears when both levels of reflection are removed.
These observations underscore the importance of both planning strategy and refining content.
Notably, the quality decreases more when removing content reflection than strategic reflection.
We speculate that strategic reflection focuses on improving aspect-level accuracy and efficiency of the entire process, providing macro-level guidance that indirectly influences the final quality.
In contrast, content reflection offers a valuable evaluation of the explanation content, serving as micro-level guidance directly related to explanation quality.
Thus, content reflection has a greater impact on the quality metrics.
By integrating these two complementary reflections, our comprehensive framework achieves superior performance.

\begin{figure}[t]
	\centering
	\setlength{\fboxrule}{0.pt}
	\setlength{\fboxsep}{0.pt}
	\fbox{
		\includegraphics[width=0.99\linewidth]{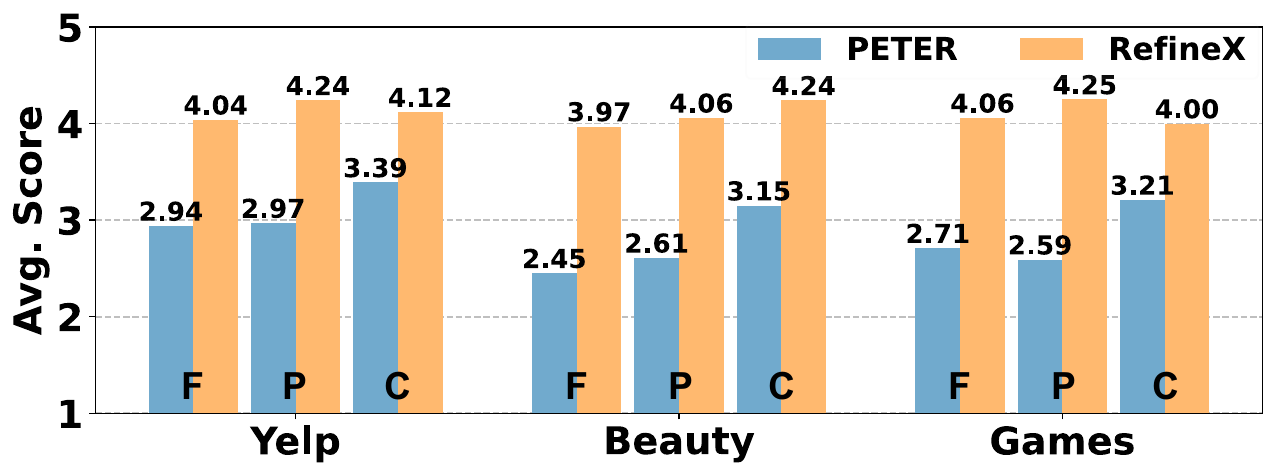}
	}
	\vspace{-0.5cm}
	\caption{Human evaluation results of PETER and RefineX on three datasets across three aspects.}
	\vspace{-0.0cm}
	\label{fig:human}
\end{figure}

\subsection{Adaptability Analysis}

Compared to existing approaches, a significant advantage of our framework is its high adaptability to various user goals. 
To verify this, following the setting of prior studies~\cite{gao2024llm}, we define three distinct user goals by assigning different weights to aspects and analyze the performance on explanation quality, refined aspect ratio, and refinement trajectory.

\noindent $\bullet$ F=P=C: Assign equal importance to three aspects: factuality, personalization and coherence.

\noindent $\bullet$ F>P>C: Assign primary importance to factuality, followed by personalization, and then coherence.

\noindent $\bullet$ P>F>C: Assign primary importance to personalization, followed by factuality, and then coherence.

The results from the Yelp and Beauty datasets are presented in Table~\ref{tab:adaptability}. 
The findings are as follows:
Our framework can flexibly tailor its refinement strategy to align with various goals, which pays more attention to important aspects and improves performance on corresponding metrics.
From the perspective of refinement trajectory, our framework prioritizes refining the most important aspects and may refine them multiple times.
This is also verified by the observed ratio of refined aspects.
Notably, when emphasizing personalization in the Yelp dataset, this aspect is often refined multiple times.
We speculate that the larger size of the Yelp dataset provides more features. 
The refinement seeks to include more features in explanations to continuously enhance personalization.

Additionally, although some refinement processes reach the maximum number of rounds (defaulted to 6), the average trajectory length remains around 4.
This demonstrates the effectiveness of our planning mechanism, which achieves precise aspect selection and efficient refinement, thereby supporting adaptability to diverse user goals.

\begin{figure*}[t]
	\centering
	\setlength{\fboxrule}{0.pt}
	\setlength{\fboxsep}{0.pt}
	\fbox{
		\includegraphics[width=0.98\linewidth]{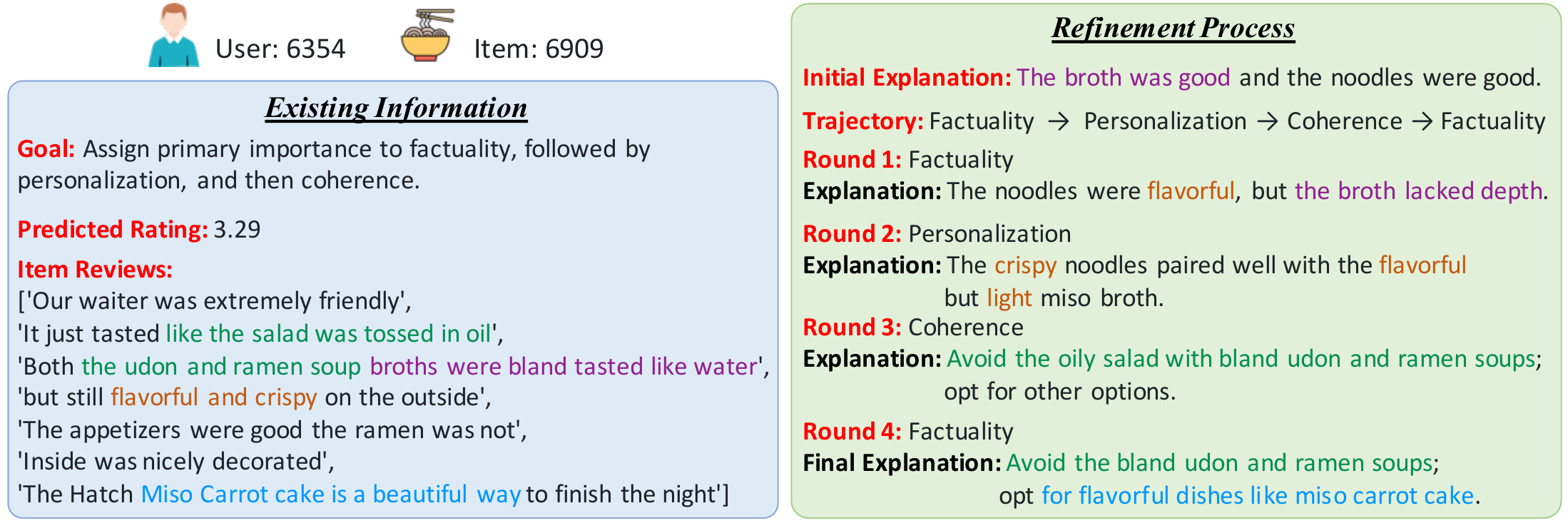}
	}
	\vspace{-0.2cm}
	\caption{An example of the refinement process. The font colors in the explanations correspond to specific content in the existing reviews.}
	\vspace{-0.5cm}
	\label{fig:case}
\end{figure*}

\subsection{Human Evaluation}
To further investigate whether the generated explanations truly assist users, we conduct a human evaluation with five experts in recommender systems. 
They are asked to evaluate explanations on three aspects: factuality, personalization, and coherence.
For each aspect, experts rate their agreement in terms of how well the explanation aligns with the corresponding standards using a 5-point Likert scale (1-strongly disagree, 2-disagree, 3-neutral, 4-agree, 5-strongly agree).
Considering efficiency and cost, we randomly select 30 pairs of explanations produced by PETER and RefineX for each dataset and present them in random order.
Thus, each expert provides a total of 540 ratings (2 methods $\times$ 30 samples $\times$ 3 aspects $\times$ 3 datasets). 

Figure~\ref{fig:human} displays the average scores for each dataset, showing that experts consistently give higher scores to explanations from our framework than those from PETER.
On average, our framework improves PETER's performance by approximately 49.0\%, 53.6\% and 26.8\% across three aspects, respectively.
These results further highlight the limitations of traditional explainable models in user-centric aspects.
Our framework enhances explanation quality through targeted refinement.
The human evaluation results show a high consistency with metrics-based evaluation in Table~\ref{tab:overall}, providing a more comprehensive assessment.

\subsection{Case Study}

To provide a more intuitive understanding of our framework, we present an example of the refinement process in Figure~\ref{fig:case}. 
The following insights can be drawn:
On one hand, our framework performs effective planning to align with the user goal. 
Specifically, the refinement trajectory closely follows the priorities specified in the user goal, where the most important aspect is refined first and multiple times.
On the other hand, our framework exhibits effective refinement capabilities.
For instance, the content ``the broth was good'' in the initial explanation conflicts with the descriptions in existing reviews.
Our framework identifies this inconsistency and corrects it in the first round.
In the second round, RefineX refines personalization by extracting unique item features to better reflect its characteristics. 
To ensure coherence between explanation sentiment and predicted user preference (neutral for a rating of 3.29), RefineX provides a balanced view by adding some disadvantages.
In the final round, RefineX enhances factual accuracy again by adding more details observed in reviews.

Furthermore, we note that the initial explanation from PETER offers a basic overview, focusing on the restaurant's food rather than service or decor.
We speculate that, based on collaborative data, PETER predicts the user has more interest in taste over other dimensions. 
Our framework further enhances this description.
By integrating explainable models with LLM-based agents, our approach achieves both accurate and user-centric explanations.

\section{Related Work}
\label{sec:related_work}

\textbf{Explainable Recommendation.}
Recommendation explanations are pivotal for improving user satisfaction and system transparency~\cite{zhang2020explainable}. 
Recently, natural language explainable recommendation has gained more attention, which generates explanations based on sequential models, such as GRU~\cite{li2020generate}, Transformer~\cite{li2021personalized}, and pre-trained language models~\cite{li2023personalized}.
To enhance explanation quality on specific aspects, some studies focus on modifying model architectures and training processes.
For example, integrating additional components to capture auxiliary information~\cite{cheng2023explainable,zhou2024enhancing},  
and employing techniques like contrastive learning~\cite{zhuang2024improving} and reinforcement learning~\cite{yang2024fine} for targeted training.
Recent studies also explore directly generating explanations by prompting LLMs~\cite{luo2023unlocking,raczynski2023problem,lei2024recexplainer}.
Unlike these methods, our work focuses on refining explanations produced by existing models in a post-hoc manner. 
While a recent study~\cite{qin2024beyond} investigates response refinement, our framework differs in task, scenario, and methodology.

\textbf{LLM-based Autonomous Agents.}
With advancements in large language models, LLM-based autonomous agents have gained significant attention~\cite{wang2024survey}, showcasing remarkable abilities in reasoning~\cite{yao2022react}, planning~\cite{shinn2024reflexion}, and tool use~\cite{schick2024toolformer}. 
Applications in this field can be divided into two categories:
The first focuses on assisting humans with complex tasks, such as software development~\cite{qian2023communicative} and role-playing in games~\cite{wang2023voyager}, while the second aims to simulate human behaviors in diverse scenarios~\cite{park2023generative,gao2023s,liu2023training}.
Several studies apply LLM-based agents to recommender systems.
Some improve recommendation performance by equipping agents with recommendation tools~\cite{huang2023recommender, zhang2024agentcf, wang2023recmind}.
Some simulate user behaviors in recommendation scenarios~\cite{zhang2024generative,wang2023user}.
In contrast to these studies, our work is the first to design an LLM-based agent framework specifically for generating recommendation explanations.

\section{Conclusion}
\label{sec:conclusion}
In this paper, we highlight the limitations of existing explainable recommender models in meeting user-centric demands and propose a novel paradigm for targeted explanation refinement.
To achieve this, we design an LLM-based multi-agent collaborative framework for explanation refinement, which employs a plan-then-refine pattern to align with user demands and incorporates a hierarchical reflection mechanism for continuous improvement. 
Extensive experiments demonstrate our framework's effectiveness in enhancing explanation quality in user-centric aspects and its high adaptability to diverse user demands, ultimately improving the helpfulness of explanations.

\section*{Limitations}
\label{sec:limitations}
This study presents the first exploration of using LLM-based agents to enhance recommendation explanations.
Despite its effectiveness, some limitations remain: 
First, the refinements in our experiments are achieved based on GPT-3.5. With the rapid development of LLMs, more advanced models could be adopted to provide more accurate refinements.
Second, we demonstrate the effectiveness of our framework in three common user-centric aspects. Other aspects, such as informativeness and comparability, could be integrated for more comprehensive refinement.
Finally, since users often have similar demands across various scenarios, our framework has potential to adapt to other generative tasks, such as dialogue generation.

\section*{Ethical Considerations}
\label{sec:ethical}
All the datasets used in our experiments are publicly available and have been widely employed in previous studies. They do not contain any personal privacy information.
Additionally, due to the training mechanisms of large language models, the generated text may contain potential biases.

\bibliography{custom}
\clearpage

\appendix
\section{Aspect Library}
\label{sec:aspect library}

To facilitate explanation refinement, we construct an aspect library containing essential materials of user-centric aspects, including aspect standards, refinement instructions, information acquisition functions, and external quality signals.
These materials support both the refinement and reflection phases.
Additionally, this structured library is designed for flexibility, enabling the dynamic combination and seamless integration of diverse aspects to accommodate personalized user goals.
Table~\ref{tab:aspect library} presents details on three aspects utilized in our experiments: Factuality, Personalization and Coherence.

\begin{table}[t]
	\centering
	\caption{{Statistics of the datasets.}}
	\vspace{-0.2cm}
	\scalebox{0.78}{
		\begin{threeparttable} 
			\begin{tabular}{p{1.0cm}<{\raggedright} p{1.1cm}<{\raggedleft} p{1.1cm}<{\raggedleft} p{1.2cm}<{\raggedleft} p{1.2cm}<{\raggedleft} p{1.5cm}<{\raggedleft}}
				\toprule
				\textbf{Dataset} & \textbf{\#User} & \textbf{\#Item} & \textbf{\#Inter.} & \textbf{Sparsity} & \textbf{Domain} \\ 
				\midrule
				\textbf{Yelp} & 15,025 & 12,445 & 698,084 & 99.63\% & Restaurant\\
				  \textbf{Beauty} & 5,396 & 3,178 & 54,805 & 99.68\% & Cosmetic\\
				\textbf{Games} & 13,957 & 7,378 & 140,353 & 99.86\% & Game\\
				\bottomrule
			\end{tabular}
		\end{threeparttable}  
	}
	\label{tab:dataset}
	\vspace{-0.0cm}
\end{table}

\begin{table*}[ht]
	\centering
	\caption{{Examples of aspect materials in the aspect library.}}
	\vspace{0.0cm}
	\small
        \setlength{\tabcolsep}{3pt}{
	\scalebox{0.9}{
		\begin{threeparttable} 
			\begin{tabular}{>{\centering\arraybackslash}m{2.5cm} >{\raggedright\arraybackslash}m{14cm}}
				\toprule
				Aspects & Materials \\ 
				\midrule
				Factuality & \textcolor{orange}{Standard:}\newline The aspect to refine is Factuality, and its standard is to Ensure the explanation is factually correct and can be supported by provided information. 
                                 \newline 
                                 \textcolor{orange}{Instruction:}\newline Refine the recommendation explanation using the information in \{Item\_Characteristics\}, ensuring the explanation is factually correct.
                                 \newline 
                                 \textcolor{orange}{Equipped Functions:}\newline get\_item\_characteristics()
                                 \newline
                                 \textcolor{orange}{Quality Signal:}\newline Entailment Ratio (Entail)
                                 \\
                    \midrule
				  Personalization & \textcolor{orange}{Standard:}\newline The aspect to refine is Personalization, and its standard is to Customize the explanation to reflect specific item characteristics and user personalities.
                                 \newline 
                                 \textcolor{orange}{Instruction:}\newline Refine the recommendation explanation using the information in \{Item\_Characteristics\} and \{User\_Personalities\}, making the explanation content personalized and reflecting user's key concerns. \newline
                                 \textcolor{orange}{Equipped Functions:}\newline get\_item\_characteristics(), get\_user\_personalities()
                                 \newline
                                 \textcolor{orange}{Quality Signal:}\newline Feature Coverage Ratio (FCR)
                                 \\
                    \midrule
				\makecell[c]{Sentiment\\Coherence} &  \textcolor{orange}{Standard:}\newline The aspect to refine is Sentiment Coherence, and its standard is to Ensure the explanation's sentiment (positive/negative) aligns with the predicted user preference (like/dislike).
                                 \newline 
                                 \textcolor{orange}{Instruction:}\newline Refine the recommendation explanation using the information in \{Item\_Pros\} and \{Item\_Cons\}. 
                                To match the explanation's sentiment with \{User\_Preference\}, emphasize advantages for positive preferences and highlight disadvantages for negative preferences.
                                 \newline 
                                 \textcolor{orange}{Equipped Functions:}\newline get\_item\_pros(), get\_item\_cons(), predict\_user\_preference()
                                 \newline
                                 \textcolor{orange}{Quality Signal:}\newline Coherence Ratio (CoR)
                                 \\
				\bottomrule
			\end{tabular}
		\end{threeparttable}}}
	\label{tab:aspect library}
	\vspace{0.0cm}
\end{table*}

\section{Details of Implementations}
\label{sec:implementation}

\subsection{Datasets}

We conduct experiments using three real-world datasets from distinct domains. Dataset statistics are presented in Table~\ref{tab:dataset}.

\subsection{Baselines} 
We provide a detailed description of each approach compared in our experiments:

$\bullet$ \textbf{PETER}~\cite{li2021personalized} is a state-of-the-art method for explainable recommendation, which personalizes the Transformer by integrating IDs with texts. We utilize it as the base model to generate initial explanations.

$\bullet$ \textbf{CLIFF}~\cite{cao2021cliff} enhances factuality in the abstractive summarization task by introducing a contrastive learning framework to distinguish positive and negative samples, subsequently extending it to the explanation generation task~\cite{zhuang2024improving}.

$\bullet$\textbf{ERRA}~\cite{cheng2023explainable} improves personalization based on PETER by incorporating an aspect enhancement component, selecting aspects most relevant to users to better capture user preference.

$\bullet$ \textbf{CER}~\cite{raczynski2023problem} aims to generate more coherent explanations based on PETER by introducing an auxiliary task of explanation-based rating estimation as a regularizer.

$\bullet$ \textbf{LLMX} employs LLMs directly to refine explanations.
Since no existing study focuses on refining explanations, we implement this method following common prompt templates from studies that generate explanations using LLMs~\cite{luo2023unlocking, lei2024recexplainer, rahdari2024logic}.

$\bullet$ \textbf{RefineX}, our proposed approach, which designs an LLM-based multi-agent collaborative refinement framework to improve explanation quality focusing on user-centric aspects.

\subsection{Evaluation Metrics} 
We utilize common metrics in the field of explainable recommendation to evaluate the quality of explanations across various aspects. 
For more precise evaluation, we use GPT-4~\footnote{gpt-4o-2024-08-06} to implement two LLM-based metrics, Entail and CoR. 
Details of each metric are as follows:

$\bullet$ \textbf{Entailment Ratio (Entail)}~\cite{xie2023factual,zhuang2024improving} measures the proportion of explanations that can be entailed or supported by existing reviews, which uses the following prompt:

\begin{tcolorbox}[colframe=black!60!white, colback=black!5!white, title={Prompt for Judging Entailment Relation.}]
You will be given a \{Recommendation\_Explanation\} and a list of existing \{Item\_Reviews\}. \newline
Your task is to evaluate whether all information in the explanation is explicitly described or implied by the reviews. \newline
- Return 1 if all information is entailed or supported by the reviews. \newline
- Return 0 if any information is not.
\end{tcolorbox}

$\bullet$ \textbf{Coherence Ratio (CoR)}~\cite{yang2021explanation, zhao2024aligning} evaluates the proportion of samples achieving coherence between explanation sentiment and predicted user preference. Sentiment is identified using the following prompt:

\begin{tcolorbox}[colframe=black!60!white, colback=black!5!white, title={Prompt for Sentiment Identification.}]
You will be given a \{Text\}, which serves as a recommendation explanation aimed to inform the user about why an item is recommended or not. \newline
Your task is to analyze the sentiment of the explanation and classify it as either positive or negative: \newline
- Positive (1): The explanation suggests recommending the item to the user. \newline
- Negative (-1): The explanation suggests not recommending the item to the user. 
\end{tcolorbox}

$\bullet$ \textbf{Feature Coverage Ratio (FCR)}~\cite{li2020generate, li2021personalized} evaluates personalization at the feature level by measuring the fraction of features present in generated explanations. It is denoted as:
\begin{equation*}
    \text{FCR} = {N}_{e} / |\mathcal{F}|,
    \label{equ: FCR}
\end{equation*}
where $N_e$ is the number of features included in the generated explanations, and $\mathcal{F}$ denotes the feature set collected in the dataset.

$\bullet$ \textbf{ENTR}~\cite{jhamtani2018learning, xie2023factual} assesses personalization from the diversity perspective by measuring the entropy of n-grams distribution in the generated text, formulated as: 
\begin{equation*}
    \text{ENTR} = \left( \textstyle \prod_{n=1}^{3} - \textstyle \sum_{x \in X_n} p(x) \log p(x)\right) ^{\frac{1}{3}},
    \label{equ: ENTR}
\end{equation*}
where each term $-\sum_{x \in X_n} p(x) \log p(x)$ represents the entropy of the unigrams, bigrams, and trigrams distribution, respectively.

\subsection{Implementation Details}
We organize each user's interactions chronologically for all datasets and divide them for training and testing using the \textit{leave-one-out} strategy~\cite{zhang2024agentcf, luo2023unlocking, sun2024large}, where the last interaction of each user is used for testing and the others are used for training.
We implement the baselines based on the code released by their authors.
For training-oriented methods, the batch size and embedding size are set to 128 and 512, respectively, and other parameters are set to their optimal values as reported in the original papers.
For RefineX, the maximum number of refinement rounds per sample is set to 6.
To ensure fair comparisons, we follow the common setting in prior studies~\cite{li2023personalized, ma2024xrec} and set the maximum length of generated explanations for all methods to 20.

\section{Prompt Design}
\label{sec:prompt}
This section introduces the prompts used by the agents within our RefineX framework, which are designed with several key components such as background clarification, system instruction, required information, and output format.
These structured prompts enable agents to execute their tasks accurately and facilitate effective collaboration, enhancing the quality of explanations on user-concerned aspects.
Detailed prompt templates are presented in Table~\ref{tab:agent prompt}.

\begin{table*}[ht]
	\centering
	\caption{{Prompt templates used in various agents.}}
	\vspace{-0.3cm}
	\small
        \setlength{\tabcolsep}{3pt}{
	\scalebox{0.9}{
		\begin{threeparttable} 
			\begin{tabular}{>{\centering\arraybackslash}m{2cm} >{\raggedright\arraybackslash}m{15cm}}
				\toprule
				Agents & Prompts \\ 
				\midrule
				Planner & 
\textcolor{orange}{\# Background Clarification} \newline
This framework refines recommendation explanations to better meet users' goals, such as Factuality, Personalization, and Sentiment coherence. \newline
It includes the following agents:\newline
- Planner: Identifies which aspect of the explanation to refine next or decides whether to terminate the process.\newline
- Refiner: Modifies the explanation on the selected aspect following the instructions.\newline
- Reflector: Evaluates the Planner's and Refiner's actions to provide feedback for improvements.\newline
Together, these agents enhance the recommendation explanation to align with user's goal.\newline

\textcolor{orange}{\# System Instruction} \newline
You are the Planner. Your role is to identify which aspect of the current explanation requires refinement in the next step, guided by the user's overall goals, refinement trajectory, and the Reflector's feedback. \newline
The framework permits up to \{Max\_Count\} modifications per explanation and will terminate when this limit is reached, necessitating careful planning of the refinement process.\newline

\textcolor{orange}{\# Required Information} \newline
The current explanation is: \{Current\_Explanation\} \newline
The user's overall goal for explanation is: \{User\_Goal\} \newline
The refinement trajectory is: \{Refinement\_Trajectory\} \newline
Reflector's feedback on the Planner's strategies: \{Strategic\_Reflection\} \newline
Reflector's feedback on the content of explanation: \{Content\_Reflection\} \newline

\textcolor{orange}{\# Output Format} \newline
\{
"aspect": <int>  \quad// Choose one: 0 (Finish), 1 (Factuality), 2 (Personalization), 3 (Sentiment Coherence)
\}
                                 \\

                    \midrule
				  Refiner & 
\textcolor{orange}{\# Background Clarification} \newline
\{{Background\_Clarification}\} \newline

\textcolor{orange}{\# System Instruction} \newline
You are the Refiner. Your role is to improve the current explanation on a specific aspect, based on the provided refinement instructions and the summarized reflections from the Reflector. \newline
Please ensure the explanation is no longer than \{Max\_Length\} words! \newline

\textcolor{orange}{\# Required Information} \newline
The current explanation is: \{Current\_Explanation\} \newline
The aspect to be refined is: \{Refined\_Aspect\} \newline
Refinement instructions and information for the refined aspect: \{Refinement\_Instruction\} \newline
Summarized Content Reflections on the refined aspect: \{Summarize\_Reflection\} \newline

\textcolor{orange}{\# Output Format} \newline
\{
    "explanation": <string> \quad// The refined explanation.
\}
                                 \\

                    \midrule
				\makecell[c]{Strategic\\Reflector} &  
\textcolor{orange}{\# Background Clarification} \newline
\{{Background\_Clarification}\} \newline

\textcolor{orange}{\# System Instruction} \newline
You are the Strategic Reflector. Your role is to evaluate the Planner's aspect-selection decisions at each round of the refinement based on the user's overall goal, refinement history and evaluation criteria. Assess whether these decisions align with the user's overall goal and provide constructive feedback to help the Planner improve.\newline

\textcolor{orange}{\# Required Information} \newline
The user's overall goal for explanation is: \{User\_Goal\} \newline 
The refinement history is: \{Refinement\_Memory\} \newline   
At the round \{Time\_Step\}, the aspect being refined is \{Refined\_Aspect\} \newline
Evaluate the Planner's selection, focusing on the following criteria: \{Strategy\_Criteria\} \newline

\textcolor{orange}{\# Output Format} \newline
\{
    "strategic reflection": <string> \quad// The generated strategic reflection.
\}
                                 \\

                    \midrule
				\makecell[c]{Content\\Reflector} &  
\textcolor{orange}{\# Background Clarification} \newline
\{{Background\_Clarification}\} \newline

\textcolor{orange}{\# System Instruction} \newline
You are the Content Reflector. Your role is to evaluate the Refiner's modifications to the content of the explanation based on the current explanation, refined aspect name, aspect instruction, quality signal and evaluation criteria. Assess whether these refinements meet the aspect standard and provide constructive suggestions for improvement.\newline

\textcolor{orange}{\# Required Information} \newline
The current explanation is: \{Current\_Explanation\} \newline
The aspect to be refined is: \{Refined\_Aspect\} \newline
Refinement instructions and information for the refined aspect: \{Refinement\_Instruction\} \newline
The external reference signal for the quality on the refined aspect is: \{Quality\_Signal\} \newline
Evaluate the Refiner's modifications to the content of explanation, focusing on the following criteria: \{Content\_Criteria\} \newline

\textcolor{orange}{\# Output Format} \newline
\{
    "content reflection": <string> \quad// The generated content reflection.
\}
                                 \\
				\bottomrule
			\end{tabular}
		\end{threeparttable}}}
	\label{tab:agent prompt}
	\vspace{0.0cm}
\end{table*}

\end{document}